\documentclass[pre,aps,twocolumn,floatfix,superscriptaddress]{revtex4}

\usepackage{graphicx}
\usepackage{dcolumn}
\usepackage{amssymb}
\usepackage{amsmath}
\usepackage{latexsym}
\usepackage{verbatim}
\usepackage{color}

\usepackage{epstopdf}
\usepackage{float}
\usepackage{epsfig}
\usepackage{enumitem}

\begin{document}

\title{Opinion formation models on a gradient}
\author{Michael T.~Gastner}
\affiliation{Department of Engineering Mathematics, University of
  Bristol, Merchant Venturers Building, Woodland Road, Bristol BS8
  1UB, UK}
\affiliation{Department of Mathematics, Imperial College London, South
  Kensington Campus, London SW7 2AZ, UK}
\affiliation{Institute of Technical Physics and Materials Science, Research Centre for Natural Sciences, Hungarian Academy of Sciences,
P.O. Box 49, H-1525 Budapest, Hungary}
\author{Nikolitsa Markou}
\affiliation{Department of Electrical and Electronic Engineering,
  Imperial College London, South Kensington Campus, London SW7 2AZ,
  UK}
\author{Gunnar Pruessner}
\affiliation{Department of Mathematics, Imperial College London, South
  Kensington Campus, London SW7 2AZ, UK}
\author{Moez Draief}
\affiliation{Department of Electrical and Electronic Engineering,
  Imperial College London, South Kensington Campus, London SW7 2AZ,
  UK}
\begin{abstract}
Statistical physicists have become interested in models of
collective social behavior such as opinion formation, where
individuals change their inherently preferred opinion if their
friends disagree.
Real preferences often depend on regional cultural differences,
which we model here as a spatial gradient $g$ in the initial
opinion.
The gradient does not only add reality to the model.
It can also reveal that opinion clusters in two dimensions are
typically in the standard (i.e.\ independent) percolation
universality class, thus settling a recent controversy about a
non-consensus model.
However, using analytical and numerical tools, we also present a model where the width of the
transition between opinions scales $\propto
g^{-1/4}$, not $\propto g^{-4/7}$ as in independent percolation,
and the cluster size distribution is consistent with
first-order percolation.
\end{abstract}
\maketitle


\section*{Introduction}
Disagreement between neighbors costs energy, in human societies as
well as in ferromagnetic spin interactions.
Because of this similarity, statistical physicists have recently shown
great interest in models of opinion formation (e.g.~\cite{Oliveira92,
  SznajdSznajd00, KrapivskyRedner03, Galam05, LambiotteRedner07,Roca_etal12},
see~\cite{Castellano_etal09, Stauffer13} for literature reviews).
Individual actors in a population are regarded as nodes in a network
and their opinions represent political affiliations, religions or
consumer choices (Microsoft Windows{\small \texttrademark} vs.\ UN*X,
Blu-ray{\small \texttrademark} vs.\ HD-DVD, etc.).
The nodes influence each other's opinions along the edges in the
network according to rules specific to the model in question.
Rules that allow a critical mass of like-minded peers to persuade a
disagreeing individual have recently found support in behavioral
experiments~\cite{Moussaid_etal13}.
The resulting opinion dynamics has been linked to election
outcomes~\cite{Bernardes_etal02, Gonzalez_etal04} and innovation
diffusion~\cite{Amini_etal09,Martins_etal09}, suggesting lessons for
political campaigns~\cite{GradowskiKosinski06} and
advertisement~\cite{WattsDodds07}.

Many opinion formation models embedded in two-dimensional space have
only one stable solution, namely complete
consensus~\cite{KrapivskyRedner03, Liggett99, LambiotteRedner07}, in
particular when they implement deterministic rules.
In reality, however, deterministic social behavior and perfect
agreement are rare~\cite{Klinkner04} -- at
least one small village of indomitable Gauls always holds out against
the Romans.
Some models thus allow clusters of a minority opinion to persist even
if entirely surrounded by the opposite opinion~\cite{Stauffer04,
  Shao_etal09}.
In this case, percolation theory provides the tools to analyze
the geometry of the minority clusters~\cite{Shao_etal09,
  Camia_etal04}.
However, the results \cite{Shao_etal09,Shao_etal12} have been subject to some controversy
because long-range correlations, thought to be responsible for
deviations from independent percolation, are expected to require a
long time to develop from an uncorrelated initial state
\cite{Sattari_etal12}.
Clearly, interactions generate complex correlations that can obscure the
familiar scaling behavior of independent 
percolation. However, 
as illustrated in
the present work, 
one must exercise great care before concluding that
 a given interaction spoils the
(asymptotic) scaling of independent percolation.

In this article we tackle the open question: can opinion dynamics, with
or without a stochastic element,
fundamentally alter percolation properties such as the clusters'
fractal dimensions or the cluster size distribution?
We show that in many cases we retrieve the scaling laws of
independent percolation.
Moreover, we also give one example where a slight change of the
dynamic rules leads to a radically different scaling behavior.

\section*{Methods}
We focus on models where the nodes are placed on a square lattice with
edges linking them to their four nearest neighbors.
Each node holds one of two possible opinions: ``black'' or
``white''.
Initially, the probability
to be black is independent at all sites and given by
\begin{equation}
  p(x) = gx+p_c,\;\;\;\;
  x\in\left[-p_c/g,(1-p_c)/g\right],
  \label{p}
\end{equation}
where $x$ is the node's horizontal position and $g\in\mathbb{R}^+$ a
constant gradient.
(We set the intercept $p_c$ equal to the percolation threshold for
later convenience.)
We interpret $p(x)$ as the innate propensity to hold the black opinion
at the beginning as well as during the evolution of the opinions.
Thus, nodes on the far left and far right of the lattice are
likely to 
have
opposite opinions.
Some previous spatial models have included heterogeneous
agents~\cite{StaufferMartins04, Centola_etal05, Mobilia_etal07}, but
no gradient.
In contrast, election results in various countries exhibit clear, smooth
gradients, especially between progressive urban and conservative rural
areas~\cite{CutlerJenkins02, ClemChodakiewicz04, Lang_etal08}.
Our model resembles such a ``culture war'' fought on a gradient.

\begin{figure}
  \begin{center}
    \includegraphics[width=8.6cm]{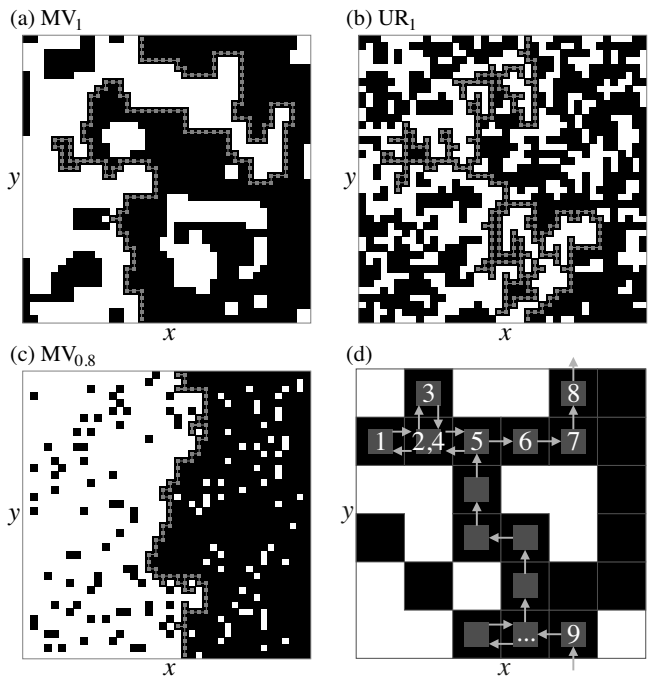}
    \caption{
      {\bf Opinion distributions and percolation hull.}
      We show typical steady-state opinion distributions for
      $g=5\times10^{-3}$ and
      (a) MV$_1$, (b) UR$_1$, (c) MV$_{0.8}$.
      The two opposing opinions are shown as black and white squares. 
      The sites marked by gray
      squares form the spanning cluster's hull. 
      (d) Illustration how the hull can be parameterized by a
      left-turning walk~\cite{GrossmanAharony86}.
    }
    \label{hull_compare}
  \end{center}
\end{figure}

Including a non-zero gradient in the numerical simulations also has
advantages for studying percolation
properties~\cite{GouyetRosso:2005}.
As opposed to running many individual simulations for a range of
different values of $p$, a gradient model allows us to analyze, in a
single simulation, clusters for a whole interval of $p$ rather than a
single fixed value.

In the present work we consider opinion formation according to the
following local rules.
\begin{itemize}
  \item Majority vote (MV): the node follows the majority opinion of
    its four nearest neighbors. If both opinions are equally represented, no
    opinion change occurs.
  \item Unanimity rule (UR): the node changes its current opinion if
    and only if all of its nearest neighbors hold the opposite
    opinion~\cite{Lambiotte_etal07}.
  \item Independent percolation (IP): the node keeps its current
    opinion irrespective of the surrounding opinions.
\end{itemize}

When a node is updated, it follows the local rule with probability $q$.
Otherwise it independently chooses a random opinion according to
Eq.~\ref{p}, so that $1-q$ is the level of noise entering the dynamics.
Notably, Eq.~\ref{p} is the only way for the local prevalence of a certain
opinion and thus the gradient to enter into the dynamics of the system.
At $q=1$ the evolution is affected by the presence of the gradient only
through the initial condition.
At $q<1$ the random updates during the evolution
exhibit the innate propensity gradient towards one or the other
opinion by allowing agents to revert to their original opinion even if
it contradicts the local majority.

All nodes simultaneously update their opinion at each time step, but
other choices such as random sequential updates do not change our
findings noticeably. The latter may have the more immediate social
interpretation as an ongoing opinion formation with agents
re-considering choices with a fixed rate, but simultaneous updates are,
surprisingly, slightly more accessible analytically. For a fixed value
of $q$, we abbreviate the models by MV$_q$ or UR$_q$, respectively.
We do not need a subscript $q$ for IP because, regardless of the value
of $q$, any snapshot of the lattice looks statistically alike,
depending only on the parameters $p_c$ and $g$ in Eq.~\ref{p}.

Once the model reaches the steady state, 
we study the geometric properties of the 
clusters formed.
On the left of Fig.~\ref{hull_compare}(a)-(c), the black clusters
form small isolated islands, whereas on the right a single large black
cluster spans from top to bottom~\cite{Sapoval_etal85}.
This percolation transition can be characterized by the hull of the
spanning cluster~\cite{Voss84}, defined as the following left-turning
walk~\cite{GrossmanAharony86, Gastner_etal09}.
We start the walk at a site with minimal $x$-coordinate in the black
spanning cluster and face towards the right
(Fig.~\ref{hull_compare}d).
First we attempt to turn to the neighbor on our left, but step in this
direction only if we reach a black site.
Otherwise, we try to move forward, then to the right, and finally
backward until we have discovered the first black neighbor.
If we iterate this procedure and apply periodic boundary conditions in
the $y$-direction, the hull has visited the entire
front of the spanning cluster when it returns to the starting
position.



\section*{Steady-state hull width and length}
If $q=1$, the dynamics is deterministic and the only source of
randomness lies in the initial assignment of opinions.
In this special case, MV$_1$ is identical to the non-consensus opinion
model of Ref.~\cite{Shao_etal09}, where it was already noted
that a small fraction of the nodes -- in our simulations 1.2\% on
average at $p_c=0.50643(1)$ -- keeps switching opinions with period
2.
When all other nodes have stopped changing opinions, we will consider
MV$_1$ to have reached its steady state.
The convergence is quick: a non-periodic node freezes after a mean of
only $0.8$ time steps.
In UR$_1$, oscillatory opinions can occur only if the initial opinions
form a perfect checkerboard pattern.
Because the gradient pins the left (right) edge to be entirely white
(black), a checkerboard pattern is impossible.
Hence, every node reaches a stationary opinion, on average after just
$0.06$ updates at $p_c=0.549199(5)$.
For IP, percolation occurs, as in zero-gradient percolation, at
$p_c=0.59274(1)$~\cite{NewmanZiff00}.

If $q<1$, the opinions in MV$_q$ and UR$_q$ never freeze, but, after a
transient, the stochastic time series of black occupancy in any column
$x$ becomes stationary.
All measurements for $q<1$ presented here were made at $q=0.8$ in this
steady state.
A visual comparison between Fig.~\ref{hull_compare}(a)-(c) suggests a
qualitative difference between MV$_1$ and UR$_1$ on the one hand and
MV$_{0.8}$ on the other hand. In the latter case, the spanning cluster
appears significantly more compact and the hull, which is centered at
$p_c=0.5000(4)$, much straighter.
So, counterintuitively, the stochastic dynamics of MV$_{0.8}$ anneals
rather than roughens the surface compared to MV$_1$ and UR$_1$.

\begin{figure}
  \begin{center}
    \includegraphics[width=7.8cm]{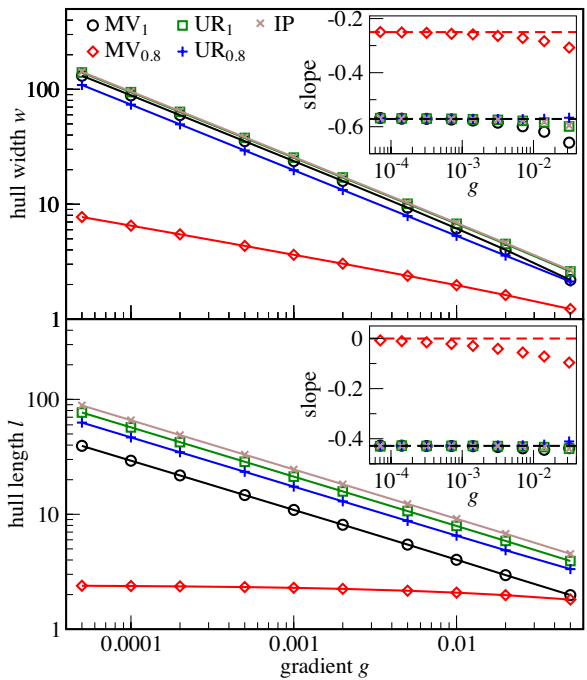}
    \caption{
      {\bf Mean hull width and length determined numerically as a
        function of the gradient.}
      Insets: slope in doubly-logarithmic scales (i.e.\
      $d\log(w)/d\log(g)$ in upper, $d\log(l)/d\log(g)$ in lower
      panel).
      Dashed lines indicate the limiting slopes for $g\to0^+$ which
      follow from scaling analysis (see text): $-4/7$
      and $-1/4$ in the upper, $-3/7$ and $0$ in the lower panel.
      Error bars are smaller than the symbol sizes.
    }
    \label{wdth}
  \end{center}
\end{figure}

We can quantify this observation by computing the hull's width $w$ and
length $l$.
If the hull consists of the walk $(x_1,y_1),\ldots,(x_l,y_l)$, we
define
\begin{equation}
  w = \sqrt{\frac{\sum_ix_i^2}l - \left(\frac{\sum_ix_i}l\right)^2}.
  \label{w}
\end{equation}
As the numerical results in Fig.~\ref{wdth} show, the width and length
for all models scale as
power laws $w\propto g^{-a}$ and $l\propto g^{-b}$ in the limit $g\to
0^+$.
With only one exception among all investigated cases, the results are
consistent with $a=4/7$ and $b=3/7$, the exact exponents of
independent gradient percolation~\cite{Nolin08}. 
We also retrieve the correlation length critical exponent $\nu$ of
standard percolation via the formula $\nu = a/(1-a) = (1-b)/b =
4/3$~\cite{Sapoval_etal85}.
The notable exception is MV$_{0.8}$ with $a=0.250(4)$ and $b=0.0074(1)$,
based on numerics for $g=10^{-4}$ and $g={5\times10^{-5}}$.
Studying the dependence of $b$ on $g$ systematically suggests $b\to0$
for $g\to0$, while $a$ stays close to $1/4$.  In
fact, the analytical results presented below indicate that $a=1/4$ and
$b=0$. In independent percolation, $a\neq4/7$ can arise only if the
probability to be black increases nonlinearly at the percolation
threshold~\cite{GastnerOborny12}.
However, in that case the ratio $b/a$ must still equal $3/4$ which is
not true for MV$_{0.8}$ so that we must look elsewhere for an
explanation.

We will briefly summarize why $a$ equals $1/4$ for MV$_q$
if $q$ is close to, but not equal to 1.
For details we refer to
Appendices 1--3.
We make two approximations. 
(1) The hull can be treated as a single-valued function of $y$ so that
we can parameterize the hull at time $t$ as a function $h(t,y)$.
(2) In MV$_{0.8}$, as opposed to UR$_q$ and IP, we observe only few
isolated minority nodes, which motivates a ``solid-on-solid''
approximation: we neglect that there is a small number of black
(white) sites to the left (right) of $h(t,y)$. 
With the notation $r=1-q$, the only transition probabilities for
$h(t,y)$ up to terms of order $O(r^2)$ are (see 
Appendix 1)
\begin{align}
  &\Pr\left[h\to h-1+K_y\right] =
  r\left[\frac12+g\left(h-\frac12+K_y\right)\right],
  \label{h_to_h-1}\\
  &\Pr\left[h\to h+K_y\right] = 
  1+r(g-1),\\
  &\Pr\left[h\to h+1+K_y\right] = 
  r\left[\frac12-g\left(h+\frac12+K_y\right)\right],
    \label{h_to_h+1}
\end{align}
where $K_y=+1$ if $h(t,y)$ is a strict local minimum in $y$, $K_y=-1$
for a maximum, and $K_y=0$ otherwise.
In the continuum limit~\cite{Vvedensky03},
the leading terms in the evolution of the hull are (see 
Appendix 2)
\begin{align}
  \frac{\partial h}{\partial t} = D\frac{\partial^2h}{\partial
    y^2} - E g h + F\eta(t,y),
  \label{dhdt}
\end{align}
where $D,E,F$ are independent of $g$ and $\eta$ is white noise with
mean zero and covariance $\langle\eta(t,y)\eta(t',y')\rangle =
\delta(t-t')\,\delta(y-y')$.
Equation \ref{dhdt} is the Edwards-Wilkinson equation
\cite{EdwardsWilkinson82} with an Ornstein-Uhlenbeck restoring force
\cite{UhlenbeckOrnstein30, vanKampen92} and can be integrated (see
Appendix 3)
to obtain the continuum limit of Eq.~\ref{w},
\begin{align}
  w^2 = \lim_{t\to\infty} \left\langle
    \overline{h(t)^2}-\overline{h(t)}^2\right\rangle =
  \frac{F^2}{4\sqrt{D E g}},
  \label{w2}
\end{align}
where the angle brackets denote the ensemble average and the overlines
symbolize spatial averages.
Thus, we obtain $w\propto g^{-1/4}$ consistent with the numerical
results for MV$_{0.8}$.
Although we have here derived the scaling law only for the MV model,
numerical evidence suggests that $a=1/4$ is valid for a broader class
of gradient models.
In Ref.~\cite{Gastner_etal11}, a numerical fit for a spatial
birth-death process on a gradient also yields $a=0.26(1)$.

\section*{Cluster sizes}

\begin{figure}
  \begin{center}
    \includegraphics[width=8.6cm]{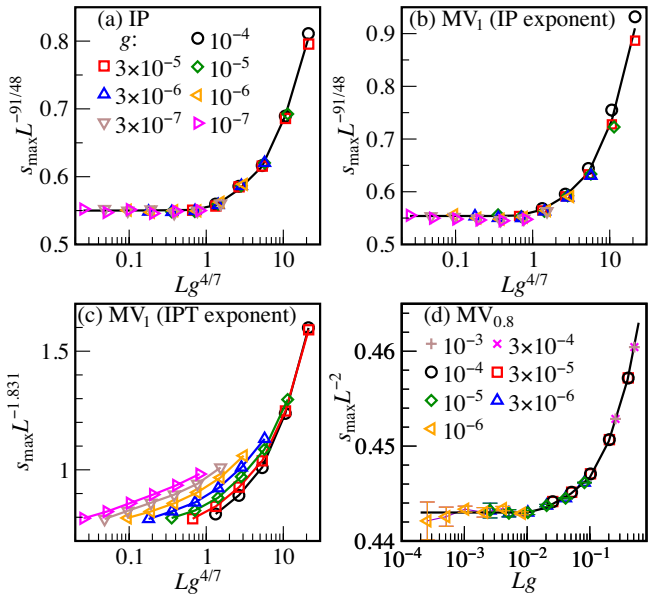}
    \caption{
      {\bf Fractal dimensions.} For the correct exponents $d_f$ and
      $c$,
      $s_{\text{max}}L^{-d_f}$ as a function of $Lg^c$ should collapse
      on a single curve with slope zero for $Lg^c\to0$.
      For (a) IP and (b) MV$_1$, $d_f=91/48$ is the same as the
      fractal dimension of standard percolation.
      (c) Replacing $d_f$ with the value $1.831$ of invasion
      percolation with trapping (IPT) does not produce a data
      collapse.
      (d) For the largest MV$_{0.8}$ cluster, we obtain a data collapse
      if $d_f=2$.
    }
    \label{smax_fig}
  \end{center}
\end{figure}

The scaling laws for $w$ and $l$ signal that MV$_{0.8}$ is not in the
same universality class as IP.
In Ref.~\cite{Shao_etal09} it is claimed that MV$_1$ is in yet
another class, namely invasion percolation with trapping (IPT).
Although $w$ scales identically in IP and IPT~\cite{Birovljev_etal91},
we now demonstrate how the gradient method can still show
unequivocally that MV$_1$ belongs to the IP class after all, thus
supporting the arguments of Ref.~\cite{Sattari_etal12}. 
We calculate the size $s_{\text{max}}$ of the largest cluster in a
lattice whose linear size is $L$ in both $x$- and $y$-direction.
We center the $x$-axis at $p_c$ so that the initial probability to
be black in Eq.~\ref{p} is limited by $\pm\frac12gL+p_c$ on the right
(left) edge.
As a function of $L$ and $g$, $s_{\text{max}}$ is expected to
satisfy the ansatz
\begin{align}
  s_{\max} = L^{d_f}f_{s_{\text{max}}}(L/\xi(g)).
  \label{smax_eq}
\end{align}
Here $d_f$ is the fractal dimension of the cluster at $p_c$,
$\xi(g)$ is the characteristic length scale for changes in the cluster
density, and the scaling function $f_{s_{\text{max}}}(z)$ approaches a
constant for $z\to0^+$.
The fractal dimensions differ between the two universality classes in
question: $d_f=91/48\approx 1.896$ for IP and $d_f=1.831(3)$ for
IPT~\cite{Schwarzer_etal99}.
Furthermore, $\xi(g)$ in IP scales linearly with $w\propto
g^{-4/7}$~\cite{Sapoval_etal85}.
Thus, according to Eq.~\ref{smax_eq}, a plot of
$s_{\text{max}}L^{-91/48}$ versus $Lg^{4/7}$ collapses the IP data for
different $L$ and $g$ on a single curve that asymptotically approaches
a constant for small $Lg^{4/7}$ (Fig.~\ref{smax_fig}a).
For MV$_1$, we obtain a data collapse with the same IP exponents
(Fig.~\ref{smax_fig}b).
By contrast, if we assume $d_f=1.831$, there is neither a collapse nor
do the individual curves approach a constant for $Lg^{4/7}\to0^+$
(Fig.~\ref{smax_fig}c),
hence ruling out that MV$_1$ is in the same universality
class as IPT. Changing the exponent $4/7$ on $g$ leads to a
lateral shift of the data in Fig.~\ref{smax_fig}(c), but we found no
value yielding a convincing data collapse. Moreover, it cannot
overcome the problem that the hypothetical scaling function
$f_{s_{\text{max}}}(z)$ would not become constant for $z\to0^+$.
However, the collapse of MV$_{0.8}$ with $d_f=2$ (which justifies the
solid-on-solid approximation in the previous section) and $\xi(g)\propto
g^{-1}$ in Fig.~\ref{smax_fig}(d) corroborates that opinion dynamics
can lead to percolation outside the IP universality class.

\begin{figure}
  \begin{center}
    \includegraphics[width=8.6cm]{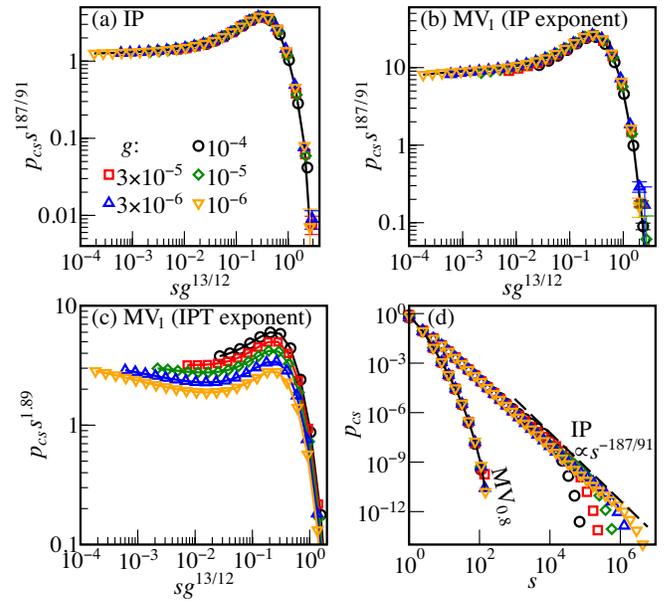}
    \caption{
      {\bf Cluster size distributions.}
      (a) The rescaled distribution $p_{cs}s^\tau$ for IP
      collapses if plotted versus $sg^{1/[\sigma(\nu+1)]}$, where
      the critical exponents $\nu$, $\sigma$, $\tau$
      are those of standard percolation.
      For MV$_1$ the data collapse is much better (b) for
      the IP exponent $\tau=187/91$ than (c) for the IPT exponent
      $\tau=1.89$.
      (d) The MV$_{0.8}$ distribution does not follow the same
      asymptotic power law as IP.
    }
    \label{clsize}
  \end{center}
\end{figure}

The cluster size distribution provides further support for this
classification.
We count all non-spanning clusters with at least one site in the
stripe $|x|<w$ and compute the fraction $p_{cs}(s)$ of clusters of
size $s$.
In IP~\cite{Gastner_etal11}
\begin{align}
  p_{cs}(s) = s^{-\tau}f_{cs}\left(sg^{1/[\sigma(\nu+1)]}\right),
  \label{p_cs}
\end{align} 
where the critical exponents are $\tau=187/91\approx2.055$,
$\sigma=36/91$, $\nu=4/3$~\cite{StaufferAharony91}, and
$f_{cs}(z)\to\text{const.}$ for $z\to0^+$ (Fig.~\ref{clsize}a).
Reference \cite{Shao_etal09} hypothesizes that in MV$_1$ the
exponent $\tau$ is replaced by $1.89(1)$, the corresponding value for
the pore size distribution in IPT.
However, Fig.~\ref{clsize}(b) and (c) show that, while the data
collapse is excellent for $\tau=187/91$, it is poor for the
alternative value $1.89$.
In summary, MV$_1$ and IP share the following critical exponents: the
hull width and length exponents $a$, $b$ and consequently $\nu=4/3$;
the fractal dimension $d_f$ and thus $\beta=\nu(2-d_f)$;
furthermore $\tau$ and $\sigma$.
This list is clear evidence that MV$_1$ is in the IP universality
class.
As shown in 
Appendix 4,
we reach the same conclusion for UR$_1$ and UR$_{0.8}$.

The situation is different in MV$_{0.8}$ where the cluster size
distribution appears to drop more sharply with a cutoff that varies
much less with the gradient.
We want to assess the lack of scaling quantitatively and distinguish
it from a power law with large exponent $\tau$ and little dependence
of the upper cutoff on $g$. 
Moment ratios $s_c^{(n)} = \langle s^{n+1}\rangle/\langle s^n\rangle$ are
asymptotically proportional to the upper cutoff, provided
$n>\tau-1$.
If the transition is continuous, then $s_c^{(n)}$ scales
asymptotically as a power of $g$.
This power law can be detected more easily than the asymptotic scaling
regime $p_{cs}\propto s^{-\tau}$~\cite{Christensen_etal08}.

\begin{figure*}
  \begin{center}
    \includegraphics[width=16.5cm]{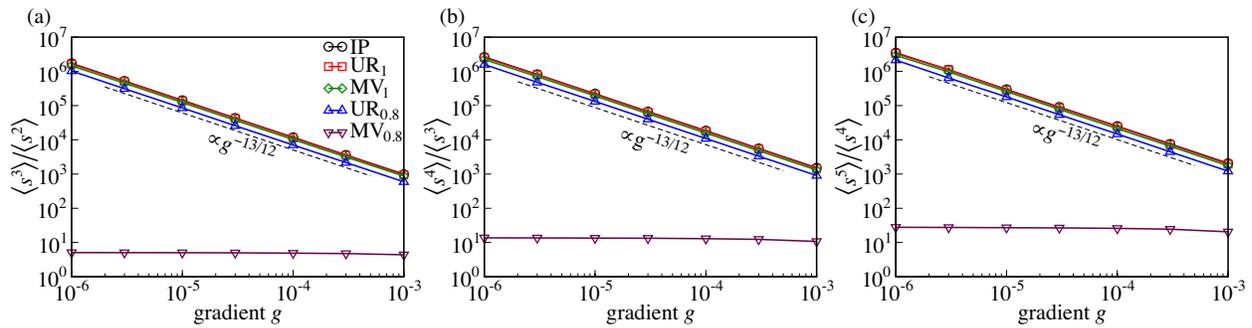}
    \caption{
      {\bf Cluster size moment ratios.}
      The moment ratios $\langle s^{n+1}\rangle/\langle s^n\rangle$ of
      the cluster size distributions for (a) $n=2$, (b) $n=3$, (c)
      $n=4$.
      The ratios for UR$_1$, MV$_1$, and UR$_{0.8}$ scale in the
      same manner as in IP, namely $\langle s^{n+1}\rangle/\langle
      s^n\rangle\propto g^{13/12}$.
      By contrast, the moment ratios for MV$_{0.8}$ appear to reach an
      asymptotic limit for $g\to0^+$.
    }
    \label{ratio_allmodels}
  \end{center}
\end{figure*}

We plot the moment ratios of IP, UR$_1$, MV$_1$, UR$_{0.8}$ and
MV$_{0.8}$ for $n=2,3,4$ in Fig.~\ref{ratio_allmodels}.
Except MV$_{0.8}$, all of these cases are in excellent agreement with
the prediction of Eq.~\ref{p_cs},
$s_c^{(n)}\propto g^{-1/[\sigma(\nu+1)]}$, where $\sigma=36/91$ and
$\nu=4/3$ are the critical exponents of IP~\cite{StaufferAharony91}.
The cutoff $s_c^{(n)}$ in MV$_{0.8}$, by contrast, does not diverge as a
power law for $g\to0^+$.
Instead $s_c^{(n)}$ appears to reach an asymptotic value for all $n$.
Such a behavior is typical of a first-order transition.
Based on these data, we can firmly rule out that $\tau$ in MV$_{0.8}$
has the IP value $187/91 \approx 2.055$.
We add the caveat that, for sufficiently large $n$, $s_c^{(n)}$ may
scale as a power of $g$ after all.
However, the data imply $\tau>5$, an unusually large value
compared to IP, directed percolation ($\tau =
2.112$)~\cite{DharBarma81} and Achlioptas percolation ($\tau =
2.04762$)~\cite{daCosta_etal10}.

\section*{Conclusion}
\begin{table*}[!ht]
\begin{tabular}{c|l|l|l}
Model & $q$ & Exponents & Universality Class \\
\hline
\parbox[t]{4cm}{Independent Percolation (IP)} &  &
$a=4/7$, $b=3/7$, $d_f=91/48$, $\nu=4/3$ & IP (by definition) \\
\hline
\parbox[t]{4cm}{Deterministic Majority\newline Vote Model (MV$_1$)} & 1 &
$a=4/7$, $b=3/7$, $d_f=91/48$, $\nu=4/3$ & IP \\
\hline
\parbox[t]{4cm}{Deterministic Unanimity\newline Rule (UR$_1$)} & 1 &
$a=4/7$, $b=3/7$, $d_f=91/48$, $\nu=4/3$ & IP \\
\hline
\parbox[t]{4cm}{Stochastic Majority\newline Vote Model (MV$_{0.8}$)} & 0.8 &
$a=1/4$, $b=0$, $d_f=2$ & Edwards-Wilkinson \\
\hline
\parbox[t]{4cm}{Stochastic Unanimity\newline Rule (UR$_{0.8}$)} & 0.8 &
$a=4/7$, $b=3/7$, $d_f=91/48$, $\nu=4/3$ & IP 
\end{tabular}
\caption{Summary of the results. For definitions of
models and exponents see text. \label{tab:summary}}
\end{table*}

We have studied in total five opinion dynamics models on a gradient, as
summarized in Table~\ref{tab:summary}. 
One of the models we studied,
independent percolation, provides the very definition of
the corresponding universality class, IP. We find that of the four other
models studied,
three display features that are fully compatible with IP,
which is commonly observed in gradient models with and without
interaction
\cite{Gouyet:1988,HaderMemsoukBoughaleb:2002,GouyetRosso:2005}.

One model, MV$_{0.8}$, differs from all of the above. 
At $p=1/2$ it has states with either a
black or white majority. Without a gradient,
(i.e.\ $g=0$ in Eq.~\ref{p}, so that $p(x)=1/2$ is constant in $x$),
there are two stable stationary solutions, where one state is above
and the other below the threshold of percolation of, say, black sites. 
There is hysteresis if one
tries to move from one majority to the other by tuning $p$, as expected for first-order transitions. By
introducing
a gradient, the two phases are
forced to collide because the left boundary must be completely white and
the right boundary black. We observe that the gradient stabilizes and
sharpens the front compared to independent percolation.

MV$_{0.8}$ differs from the other models in two important points. 
First, its stochastic nature helps anneal boundaries between opposite
opinions. 
The second difference is that the majority rule makes small clusters
more prone to invasion by the opposing opinion. The combination of these two features results in what appears to be a
first-order transition. Nevertheless, the opinion interface displays scaling,
found to be in the Edwards-Wilkinson universality class, which differs
significantly from independent percolation.

The birth-death model of Ref.~\cite{Gastner_etal11}
suggested already the possibility of first-order transitions in gradient
models. We leave it to future research to analytically confirm the
first-order nature of the MV$_{0.8}$ transition. It would also be insightful
to investigate more complex network topologies that are based on real
social interactions rather than a regular square lattice. We emphasize
that, in the light of previous work on explosive percolation
\cite{Achlioptas_etal09,daCosta_etal10, RiordanWarnke11,Cho_etal13}, only
analytic results can fully clarify the order of any percolation
transition. However, we can conclude with certainty that, although none
of the opinion models we have investigated is consistent with IPT,
MV$_{0.8}$
is an example of a dynamic rule that leads to percolation outside the IP
universality class.

From a sociological perspective, our study shows that 
small variations in the innate propensity towards one or another
opinion may turn into a spatial discontinuity in the opinions.
Interestingly, the sharpest division occurs when agents do not follow the
local majority all the time.
Hence, processes 
that may 
be perceived as having the effect of making the interface between
different opinions more blurred, such as the majority rule 
with
stochasticity involved, have the opposite effect.
They anneal this
interface and contribute to the collapse of minority clusters, which
are sustained in the presence of stricter rules, such as the
deterministic unanimity rule.

\section*{Acknowledgments}
MTG acknowledges support by Imperial College London and the European
Commission (project number FP7-PEOPLE-2012-IEF 6-4564/2013).

\bibliography{opin_form_grad}

\section*{Appendix 1: Transition probabilities in the hull dynamics of
  the majority vote model (Eq.~\ref{h_to_h-1}--\ref{h_to_h+1})}

Let the hull be parameterized as $(x_1,y_1),\ldots,(x_l,y_l)$ by the
left-turning walk described above at the end of the Methods section.
Figure~\ref{hull_compare}c shows that the hull in MV$_{0.8}$ separates
a predominantly white region on the left from a similarly dense black
region on the right.
This observation justifies a ``solid-on-solid''
approximation~\cite{Kroll81} where we ignore
\begin{itemize}
\item any overhangs in the interface (i.e., parts of the left-turning
  walk that move towards smaller $y$-coordinates), 
\item any isolated islands of the minority
  color to the left and right of the hull.
\end{itemize}
In this approximation, the hull at time step $t$ is completely
characterized by 
\begin{align}
  h(t,y) = \min\{x_k|y_k=y,k=1,\ldots,l\}-\frac12
\end{align}
because every site $(x,y)$ with $x<h(t,y)$ will be white and every
site with $x>h(t,y)$ black.
We now have to distinguish three cases.

\begin{figure}
  \begin{center}
    \includegraphics[width=8.5cm]{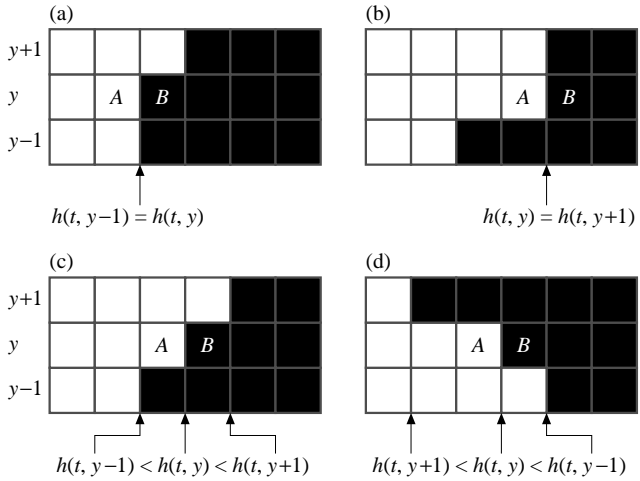}
    \caption{
      Examples where $h(t,y)$ is neither a strict local
      minimum nor maximum.
      In each case, the white and black sites at the interface, $A$ and
      $B$, have at least two neighbors of their own color so that in
      MV$_1$ there is no change of the hull position (i.e., $h(t,y)
      = h(t+1,y)$).
      In MV$_{1-r}$, there is a $O(r)$ probability that the hull moves
      one site to the left or right.
      All other probabilities are $O(r^2)$.
    }
    \label{no_strict_minmax}
  \end{center}
\end{figure}

\subsubsection*{Case 1: $h(t,y)$ is neither a strict local minimum nor
  maximum}
In this case, one of the following four conditions must be
met 
\begin{itemize}
\item $h(t,y-1)=h(t,y)$,
\item $h(t,y)=h(t,y+1)$,
\item $h(t,y-1) < h(t,y) < h(t,y+1)$, or 
\item $h(t,y-1) > h(t,y) > h(t,y+1)$.
\end{itemize}
Let $A$ be the white site in the $y$-th row with $x$-coordinate
$h(t,y)-\frac12$ and $B$ the black site at $h(t,y)+\frac12$ (see
Fig.~\ref{no_strict_minmax}).
In all cases listed above, both $A$ and $B$ have at least two
neighbors of their own color, namely one in the $y$-th row and one in a
neighboring row.
Including their own vote, the local majority supports their current
opinion.
As a consequence, in the deterministic majority vote model MV$_1$
neither $A$ nor $B$ will change color and thus
$h(t+1,y)=h(t,y)$.
In the stochastic model MV$_{1-r}$ with $r>0$, the probability that 
$A$ becomes black is 
\begin{align}
  \Pr[\text{$A$ black at $t+1$}\,|\,h(t,y)=x] =\nonumber\\
  r\left[g\left(x-\frac12\right)+p_c\right],
\end{align}
and the probability that $B$ becomes white
\begin{align}
  \Pr[\text{$B$ white at $t+1$}\,|\,h(t,y)=x] =\nonumber\\
  r\left[1-g\left(x+\frac12\right)-p_c\right].
\end{align}
In the solid-on-solid approximation, the hull can shift exactly one
step to the left only if $A$ turns black, while all other sites keep
their colors with a probability $1-O(r)$.
Because the probabilities are independent, we can multiply them and
obtain
\begin{align}
  \Pr\left[h(t+1,y)=x-1\,|\,h(t,y)=x\right] =\nonumber\\
  r\left[g\left(x-\frac12\right)+p_c\right] + O(r^2).
\end{align}
Similarly,
\begin{align}
  \Pr[h(t+1,y)=x+1\,|\,h(t,y)=x] = \nonumber\\
  r\left[1-g\left(x+\frac12\right)-p_c\right] + O(r^2).
\end{align}
Because shifts of more than one step to either side have probabilities
$O(r^2)$ and the sum of the probabilities must equal one,
\begin{align}
  \Pr[h(t+1,y)=x\,|\,h(t,y)=x] = \nonumber\\
  1 + r(g-1) + O(r^2).
\end{align}

\begin{figure}
  \begin{center}
    \includegraphics[width=9cm]{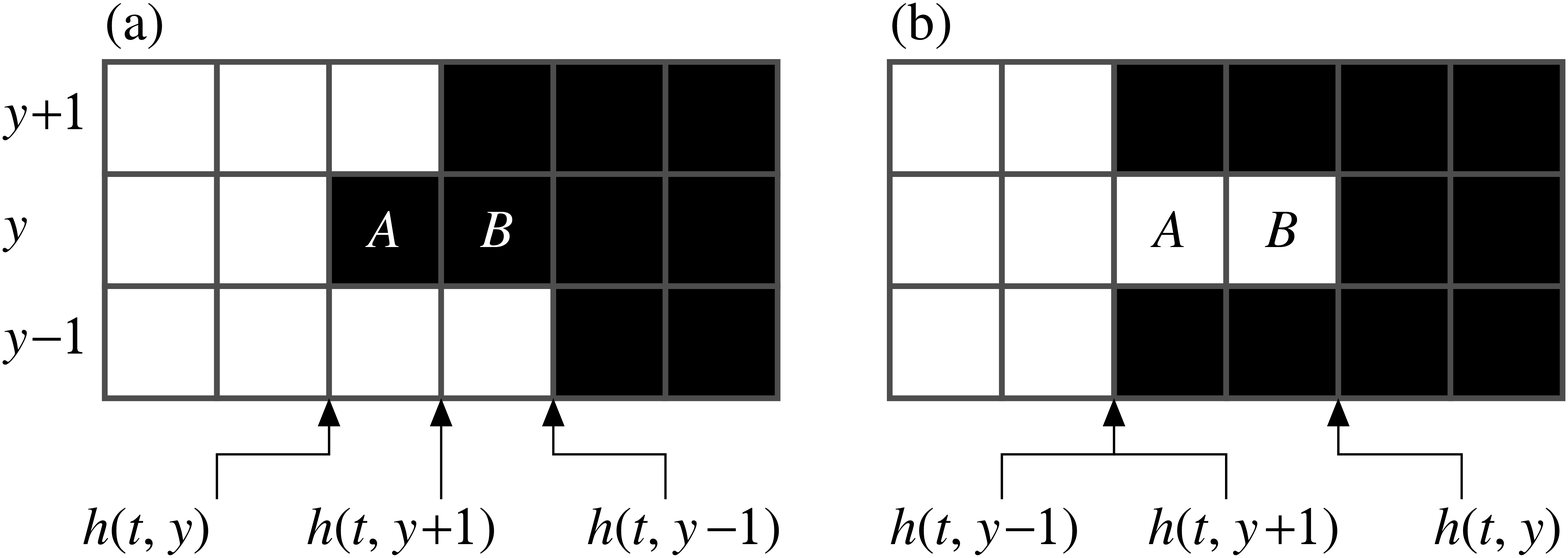}
    \caption{
      Examples where $h(t,y)$ is a strict local (a) minimum or (b)
      maximum.
      The minimal or maximal site of the protruding opinion is in a
      local minority, but all other sites in row $y$ will have at
      least two neighbors of the same opinion.
      Thus, in MV$_1$ only the front site will change opinion between
      time steps $t$ and $t+1$.
      In MV$_{1-r}$ the probability of the hull shifting one step
      towards the (a) right, (b) left is $1-O(r)$.
      The probability of no change or two steps to the (a) right, (b)
      left is $O(r)$.
      All other transitions have probabilities $O(r^2)$.
    }
    \label{strict_minmax}
  \end{center}
\end{figure}

\subsubsection*{Case 2: $h(t,y)$ is a strict local minimum}
Here the leftmost black site $A$ in row $y$ is in a local minority
(see Fig.~\ref{strict_minmax}a).
It stays black with probability
\begin{align}
  \Pr[\text{$A$ black at $t+1$}\,|\,h(t,y) = x] = \nonumber\\
  r\left[g\left(x+\frac12\right)+p_c\right].
\end{align}
Its right neighbor $B$ is black, which is the local majority because
at least its two neighbors in row $i$ are black. 
Hence, it becomes white with probability
\begin{align}
  \Pr[\text{$B$ white at $t+1$}\,|\,h(t,y)=x] = \nonumber\\
    r\left[1-g\left(x+\frac32\right)-p_c \right].
    \label{B_white}
\end{align}
In the solid-on-solid approximation, the hull can only stay at the
same position if none of the sites in row $i$ changes its color. With the
only exception of $A$, an individual opinion change has probability
$1-O(r)$ for all sites so that
\begin{align}
  \Pr[h(t+1,y)=x\,|\,h(t,y)=x] = \nonumber\\
  r\left[g\left(x+\frac12\right)+p_c\right] + O(r^2).
\end{align}
The hull can only shift two sites to the right if $A$ and $B$ become
white.
The former has probability $1-O(r)$, the latter is given by Eq.\
\ref{B_white}, and all other probabilities are $1-O(r)$.
Therefore,
\begin{align}
  \Pr[h(t+1,y)=x+2\,|\,h(t,y)=x] = \nonumber\\
  r\left[1-g\left(x+\frac32\right)-p_c \right] + O(r^2).
\end{align}
All other shifts further to the left and right are $O(r^2)$, so
that the only remaining transition of one step to the right has
probability
\begin{align}
  \Pr[h(t+1,y)=x+1\,|\,h(t,y)=x] =\nonumber\\
  1+r(g-1) + O(r^2).
\end{align}

\subsubsection*{Case 3: $h(t,y)$ is a strict local maximum}
In analogy to case 2, we find
\begin{align}
  &\Pr[h(t+1,y)=x\,|\,h(t,y)=x] =\nonumber\\
  &\text{\hspace{1cm}}r\left[1-g\left(x-\frac12\right)-p_c\right]+O(r^2),\\
  &\Pr[h(t+1,y)=x-2\,|\,h(t,y)=x] =\nonumber\\
  &\text{\hspace{1cm}}r\left[g\left(x-\frac32\right)+p_c\right]+O(r^2),\\
  &\Pr[h(t+1,y)=x-1\,|\,h(t,y)=x] =\nonumber\\
  &\text{\hspace{1cm}}1+r(g-1)+O(r^2).
\end{align}

\subsubsection*{Summary}
We can summarize the results so far with the notation
\begin{align}
  K_y =
  \begin{cases}
    +1&\text{ if $h(t,y)$ is a strict minimum,}\\
    -1&\text{ if $h(t,y)$ is a strict maximum,}\\
    0&\text{ otherwise}.
  \end{cases}
  \label{K_y}
\end{align}
Neglecting terms $O(r^2)$,
\begin{align}
  &\Pr\left[h(t+1,y)=x-1+K_y\,|\,h(t,y)=x\right] =\nonumber\\
  &\text{\hspace{1cm}}r\left[g\left(x-\frac12\right)+p_c\right] + rgK_y,
  \label{x2x-1}\\
  &\Pr\left[h(t+1,y)=x+K_y\,|\,h(t,y)=x\right] =\nonumber\\
  &\text{\hspace{1cm}}1+r(g-1),\\
  &\Pr\left[h(t+1,y)=x+1+K_y\,|\,h(t,y)=x\right] =\nonumber\\
  &\text{\hspace{1cm}}r\left[1-g\left(x+\frac12\right)-p_c\right] - rgK_y.
  \label{x2x+1}
\end{align}
Because there are no isolated clusters in the solid-on-solid
approximation and the dynamics is symmetric under interchange of black
and white, $p_c$ equals $\frac12$.
This assumption is consistent with our numerical results for the full
model $p_c=0.5000(4)$.
Equations \ref{x2x-1}--\ref{x2x+1} with $p_c=\frac12$ yield Eq.~\ref{h_to_h-1}--\ref{h_to_h+1}.

\section*{Appendix 2: The stochastic differential equation for the hull evolution
  (Eq.~\ref{dhdt})}

An alternative formulation of Eq.~\ref{x2x-1}--\ref{x2x+1} is
\begin{align}
  h(t+1,y) = h(t,y) + K_y + \zeta_y,
  \label{difference_eq}
\end{align}
where
\begin{align}
  &\Pr(\zeta_y=-1) =
  r\left[\frac12+g\left(h(t,y)-\frac12+K_y\right)\right],\\
  &\Pr(\zeta_y=0) = 1+r(g-1),\\
  &\Pr(\zeta_y=1) =
  r\left[\frac12-g\left(h(t,y)+\frac12+K_y\right)\right].
\end{align}
The expectation value of $\zeta_y$ is
\begin{align}
  \langle\zeta_y\rangle = -2gr(h+K_y),
\end{align}
so that we can rephrase Eq.~\ref{difference_eq} as
\begin{align}
  \underbrace{h(t+1,y) - h(t,y)}_A = K_y - \underbrace{2gr(h+K_y)}_B +
  \underbrace{\zeta_y - \langle\zeta_y\rangle}_C.
  \label{difference_eq_rewritten}
\end{align}
Our objective is to take the continuum limit of Eq.\
\ref{difference_eq_rewritten} in the following manner.
With the notation
\begin{align}
  \Delta_+ = h(t,y+1)-h(t,y),\\
  \Delta_- = h(t,y)-h(t,y-1),
\end{align}
we can express $K_y$ of Eq.~\ref{K_y} using the Heaviside
step function
\begin{align}
  \theta(x) = 
  \begin{cases}
    1&\text{ if $x\geq0$},\\
    0&\text{ otherwise}
  \end{cases}
\end{align}
as
\begin{align}
  &K_y = \left[1-\theta\left(-\Delta_+\right)\right]
  \left[1-\theta\left(\Delta_-\right)\right]\nonumber\\
  &\text{\hspace{1cm}}-
  \left[1-\theta\left(\Delta_+\right)\right]
  \left[1-\theta\left(-\Delta_-\right)\right].
  \label{K_y2}
\end{align}
The discontinuous Heaviside function can be written as the limit
$\epsilon\to0$ of the differentiable function~\cite{Vvedensky03}
\begin{align}
  \theta_\epsilon(x) = \epsilon\,\ln\left(
    \frac{\exp\left(\frac{x+1}\epsilon\right)+1}
    {\exp\left(\frac x\epsilon\right)+1}\right).
  \label{theta_epsilon}
\end{align}
Simultaneously with the limit of the Heaviside function, we take the
continuum limit of the space and time variables,
\begin{align}
  &\tilde{t} = \epsilon^k t,\\
  &\tilde{y} = \epsilon^l y,\\
  &\tilde{h}(\tilde{t},\tilde{y}) = \epsilon^m h(t,y),\label{h_tilde}
\end{align}
with $k,l,m>0$
and let $g$ approach zero as
\begin{align}
  g = \epsilon^n\tilde{g}
  \label{r_limit}
\end{align}
with $n>0$.
We will now determine the leading terms in the individual parts of
Eq.~\ref{difference_eq_rewritten}, which will give us conditions for
these exponents.

\subsubsection*{I: The discrete time derivative $A$ in
  Eq.~\ref{difference_eq_rewritten}}
Assuming that $h$ is a smooth function, we can expand $A$ as
\begin{align}
  A &=
  \epsilon^{-m}\left(\tilde{h}(\tilde{t}+\epsilon^k) +
    \tilde{h}(\tilde{t})\right) \nonumber\\
&=
  \epsilon^{k-m}\,
    \frac{\partial\tilde{h}}{\partial\tilde{t}} +
    O\left(\epsilon^{2k-m}\right).
\end{align}

\subsubsection*{II: The variable $K_y$ encoding a strict minimum or maximum}
For the derivatives of $\theta_\epsilon$ of Eq.~\ref{theta_epsilon},
we find
\begin{align}
  \theta_\epsilon(0) &= 1-\epsilon\,\ln(2) +
  O\left(\epsilon\,e^{-1/\epsilon}\right),
\end{align}
\begin{align}
  \frac{d\theta_\epsilon(0)}{dx} &=
  \frac12+O\left(e^{-1/\epsilon}\right),
\end{align}
\begin{align}
  \frac{d^2\theta_\epsilon(0)}{dx^2} &= -\frac 1{4\epsilon} +
  O\left(\epsilon^{-1}e^{-1/\epsilon}\right),
\end{align}
\begin{align}
  \frac{d^3\theta_\epsilon(0)}{dx^3} &=
  O\left(\epsilon^{-2}e^{-1/\epsilon}\right).
\end{align}
Inserting these derivatives into Eq.~\ref{K_y2} we obtain
\begin{widetext}
\begin{align}
  K_{y,\epsilon} &= \left[\epsilon\,\ln(2)+\frac12\Delta_+ -
    \frac1{8\epsilon}\Delta_+^2 +
    O\left(\epsilon^{-3}\Delta_+^4\right)\right]
  \left[\epsilon\,\ln(2)-\frac12\Delta_- -
    \frac1{8\epsilon}\Delta_-^2 +
    O\left(\epsilon^{-3}\Delta_-^4\right)\right]\nonumber\\
  &\hspace{0.5cm}-\left[\epsilon\,\ln(2)+\frac12\Delta_- -
    \frac1{8\epsilon}\Delta_-^2 +
    O\left(\epsilon^{-3}\Delta_-^4\right)\right]
  \left[\epsilon\,\ln(2)-\frac12\Delta_+ -
    \frac1{8\epsilon}\Delta_+^2 +
    O\left(\epsilon^{-3}\Delta_+^4\right)\right] \nonumber\\
  & = \epsilon\,\ln(2)\left(\Delta_+-\Delta_-\right) +
  \frac1{8\epsilon}\Delta_+\Delta_-\left(\Delta_+-\Delta_-\right) +
  O\left(\epsilon^{-3}\Delta^5\right).
\end{align}
\end{widetext}
From the Taylor expansions of $\tilde{h}$ we obtain
\begin{align}
  &\Delta_+-\Delta_- =
  \epsilon^{2l-m}\frac{\partial^2\tilde{h}}{\partial\tilde{y}^2} +
  O\left(\epsilon^{4l-m}\right),
\end{align}
\begin{align}
  &\Delta_+\Delta_-\left(\Delta_+-\Delta_-\right) =\nonumber\\
  &\text{\hspace{1cm}}\epsilon^{4l-3m}
  \left(\frac{\partial\tilde{h}}{\partial\tilde{y}}\right)^2
  \frac{\partial^2\tilde{h}}{\partial\tilde{y}^2} +
  O\left(\epsilon^{6l-3m}\right),
\end{align}
so that
\begin{align}
  K_{y,\epsilon} =
  \epsilon^{2l-m+1}\ln(2)\frac{\partial^2\tilde{h}}{\partial\tilde{y}^2}
  + O\left(\epsilon^{4l-3m-1}\right),
  \label{Ky_leading}
\end{align}
where the expansion converges only if 
\begin{align}
  l-m>1.
  \label{l_minus_m}
\end{align}

\subsubsection*{III: The gradient term $B$ in Eq.~\ref{difference_eq_rewritten}}
From Eq.~\ref{h_tilde}, \ref{r_limit} and \ref{Ky_leading},
we obtain
\begin{align}
  B &=
  2\epsilon^n\tilde{g}r
  \left[\epsilon^{-m}\tilde{h}(\tilde{t},\tilde{y}) +
    O\left(\epsilon^{2l-m+1}\right)\right]\nonumber\\
  &= 2\epsilon^{n-m}\tilde{g}r\tilde{h}(\tilde{t},\tilde{y}) +
  O\left(\epsilon^{2l-m+n+1}\right).
\end{align} 

\subsubsection*{IV: The noise term $C$ in Eq.~\ref{difference_eq_rewritten}}
The covariance of the noise is 
\begin{align}
  \Big\langle\left[\zeta_y(t)-\langle\zeta_y(t)\rangle\right]\times
    \left[\zeta_{y'}(t')-\langle\zeta_{y'}(t')\rangle\right]\Big\rangle=
  \nonumber\\
  \left[r(1-g)+O\left(r^2\right)\right]\delta_{t,t'}\delta_{y,y'}.
\end{align}
In the continuum limit, the Kronecker deltas transform as
\begin{align}
  &\delta_{t,t'} = \epsilon^k\delta(\tilde{t}-\tilde{t}'),\\
  &\delta_{y,y'} = \epsilon^l\delta(\tilde{y}-\tilde{y}').
\end{align}
Dropping terms of order $O(r^2)$,
\begin{align}
  &\langle C(\tilde{t},\tilde{y})C(\tilde{t}',\tilde{y}')\rangle
  =\nonumber\\
  &\text{\hspace{1cm}}\epsilon^{k+l}
  r\delta(\tilde{t}-\tilde{t}')\delta(\tilde{y}-\tilde{y}') +
  O\left(\epsilon^{k+l+n}\right),
\end{align}
where $C(\tilde{t},\tilde{y})=\zeta_{\tilde{y}}(\tilde{t}) -
\left\langle \zeta_{\tilde{y}}(\tilde{t}) \right\rangle$ is defined as
in Eq.~\ref{difference_eq_rewritten}.
If we introduce the rescaled noise
\begin{align}
  \eta(\tilde{t},\tilde{y}) =
  \frac{\epsilon^{-(k+l)/2}}{\sqrt{r}}C(\tilde{t},\tilde{y}),
\end{align}
then the covariance
\begin{align}
  \langle\eta(\tilde{t},\tilde{y})\eta(\tilde{t}',\tilde{y}')\rangle =
  \delta(\tilde{t}-\tilde{t}')\delta(\tilde{y}-\tilde{y}') +
  O(\epsilon^n)
\end{align}
is to highest order independent of $\epsilon$.

\subsubsection*{V: Summary}
Including only the leading terms, Eq.~\ref{difference_eq_rewritten}
becomes 
\begin{align}
  \epsilon^{k-m}\frac{\partial\tilde{h}}{\partial\tilde{t}} = &
  \epsilon^{2l-m+1}\ln(2)\frac{\partial^2\tilde{h}}{\partial\tilde{y}^2}
  - 2\epsilon^{n-m}\tilde{g}r\tilde{h}(\tilde{t},\tilde{y}) \nonumber\\&+ 
  \sqrt{r}\epsilon^{(k+l)/2}\eta(\tilde{t},\tilde{y}).
  \label{epsilon_differential_eq}
\end{align}
The four different terms scale with the same power of $\epsilon$ if
\begin{align}
  &l = \frac12(k-1),\\
  &m = \frac14(k+1),\\
  &n = k.
\end{align}
The inequality of Eq.~\ref{l_minus_m} can be satisfied by $k>7$.
Dividing Eq.~\ref{epsilon_differential_eq} by $\epsilon^{k-m} =
\epsilon^{2l-m+1} = \epsilon^{n-m} = \epsilon^{(k+l)/2}$
yields Eq.~\ref{dhdt}.

\section*{Appendix 3: The derivation of the hull width (Eq.~\ref{w2})}

We consider Eq.~\ref{dhdt}
with periodic boundaries at $y=0$ and $y=L$.
The solution $G(t,y)$ of the deterministic equation
\begin{equation}
  \frac{\partial G}{\partial t} = D\frac{\partial^2G}{\partial
    y^2} - E g G
\end{equation}
with initial condition $\lim_{t\to 0}G(t,y) = \delta(y-y_0)$ is
\begin{align}
  &G(t,y;y_0) = \nonumber\\
  &\text{\hspace{0.5cm}}\frac1L\sum_{n=-\infty}^\infty\exp\left[-(D
    k_n^2+E g)t + ik_n(y-y_0)\right],
\end{align}
where $k_n = 2\pi n/L$.
The hull position can be expressed in terms of $G$ as
\begin{align}
  h(t,y) = F\int_0^Ldy_0\int_0^tdt'G(t-t',y;y_0)\eta(t',y_0).
\end{align}
Combining the last two expression, we can derive
\begin{widetext}
\begin{align}
  \langle h(t,y)h(t,y')\rangle =
  \frac{F^2}{2L}\sum_{n=-\infty}^\infty\frac
  {1-\exp\left(-2\left(D k_n^2+E g\right)t\right)}
  {D k_n^2+E g}\exp\left[i k_n\left(y'-y\right)\right]
\end{align}
The hull width is
\begin{align}
  w^2(L) &=
  \lim_{t\to\infty}\left\langle\frac1L\int_0^Ldy\,h^2(t,y) -
    \left(\frac1L\int_0^Ldy\,h(t,y)\right)^2\right\rangle
  = \frac{F^2}L\sum_{n=1}^\infty\left(D k_n^2+E
    g\right)^{-1}\nonumber\\
  &= \frac{F^2}2\left(\frac{\coth\left(\frac L2\sqrt{\frac{E
            g}D}\right)}{2\sqrt{D E g}} - \frac1{E
      gL} \right).
\end{align}
\end{widetext}
In the limit of large system size
\begin{align} 
  \lim_{L\to\infty}w^2(L) = \frac{F^2}{4\sqrt{DE g}},
\end{align}
which is identical with Eq.~7.

\section*{Appendix 4: UR$_1$ and UR$_{0.8}$ are in the IP universality class}

\begin{figure}
  \begin{center}
    \includegraphics[width=\linewidth]{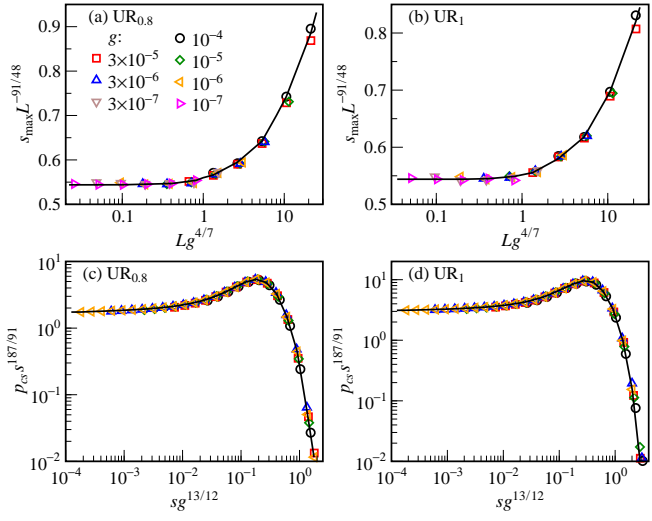}
    \caption{
      Collapse plots for (a) the size of the largest cluster
      $s_\text{max}$ in UR$_{0.8}$, (b) in UR$_1$, (c) the cluster
      size distribution $p_{cs}$ for UR$_{0.8}$, (d) UR$_1$.
    }
    \label{UR_scal}
  \end{center}
\end{figure}

In Fig.~\ref{smax_fig} and \ref{clsize} we show collapse plots for the
maximum cluster size $s_\text{max}$ and the cluster size distribution
$p_{cs}$ for IP and MV$_1$.
In these cases, the data points lie on a single curve when we plot
$s_\text{max}L^{d_f}$ versus $Lg^{\nu/(\nu+1)}$ and $p_{cs}s^{-\tau}$
versus $sg^{1/[\sigma(\nu+1)]}$.
The crucial observation is that the collapse occurs when inserting the
IP critical exponents $d_f=91/48$, $\nu=4/3$, $\tau=187/91$ and
$\sigma=36/91$ in these expressions, a telltale sign that MV$_1$ is
indeed in the IP universality class.

In Fig.~\ref{UR_scal} we make the equivalent plots
for UR$_{0.8}$ and UR$_1$ with the same exponents.
The data again fall on a single curve in each case.
Moreover, we have already seen in Fig.~\ref{wdth} that for
both of these models $w\propto g^{-\nu/(\nu+1)}$ and $b\propto
g^{-1/(\nu+1)}$ as in IP.
Thus, all numerical evidence points to both UR$_{0.8}$ and
UR$_1$ (unlike MV$_{0.8}$) belonging to the IP universality class.

\end{document}